\documentclass[12pt]{article}
\usepackage{epsfig,amssymb}

\makeatletter
\def\eqalign#1{\null\vcenter{\def\\{\cr}\openup\jot\m@th
  \ialign{\strut$\displaystyle{##}$\hfil&$\displaystyle{{}##}$\hfil
      \crcr#1\crcr}}\,}
\makeatother

\newcommand{\be}{\begin{equation}} 
\newcommand{\ee}{\end{equation}}
\newcommand{\beqa}{\begin{eqnarray}}
\newcommand{\eeqa}{\end{eqnarray}}
\newcommand{\bt}{\begin{theorem}}
\newcommand{\et}{\end{theorem}}
\newcommand{\bl}{\begin{lemma}}
\newcommand{\el}{\end{lemma}}
\newcommand{\bc}{\begin{corollary}}
\newcommand{\ec}{\end{corollary}}
\newcommand{\ba}{\begin{array}}
\newcommand{\ea}{\end{array}}

\newcommand{\la}{\label}
\newcommand{\ci}{\cite}

\newcommand{\bi}{\bibitem}

\newtheorem{theorem}{Theorem}

\newtheorem{corollary}{Corollary}
\newtheorem{lemma}{Lemma}

\newcommand{\wt}{\widetilde}
\newcommand{\de}{\delta}

\newcommand{\al}{\alpha}

\newcommand{\ze}{\zeta}
\newcommand{\ka}{\varkappa}
\renewcommand{\th}{\theta}

\newcommand{\Ai}{{\mathrm{Ai}}\,}

\begin{document}
\begin{center}
{\Large\bf 
Large gap asymptotics for random matrices
}
\end{center}
\bigskip\bigskip

\centerline{ I.~Krasovsky}
\bigskip 
\bigskip\bigskip

Let $K^{(j)}_s$, $j=1,2$ be the trace-class operators with kernels
\be
K^{(1)}_s(x,y) =\frac{\sin(x-y)}{\pi(x-y)},\qquad
K^{(2)}_s(x,y) ={\Ai(x)\Ai'(y)-\Ai(y)\Ai'(x)\over x-y}
\ee
acting on $L^2(0,2s)$ and $L^2(-s,\infty)$, respectively.
We are interested in the behaviour of the following Fredholm determinants,
the so called sine-kernel and Airy-kernel determinants,
\be
P^{(j)}_s=\det(I-K_s^{(j)}),\qquad j=1,2,
\ee
as $s\to+\infty$.
In the Gaussian Unitary Ensemble of random matrices \ci{M},
$P^{(1)}_s$ is the probability, in the bulk scaling limit, that 
there are no eigenvalues in the interval $(0,2s)$; while
$P^{(2)}_s$ is the probability, in the edge scaling limit,
that there are no eigenvalues in the interval $(-s,+\infty)$
($P^{(2)}_s$ is the distribution of the largest 
eigenvalue).
The asymptotics of $P^{(j)}_s$ as $s\to+\infty$ are often referred 
to as the large gap asymptotics. 

We will describe the main steps of the method of computing the 
asymptotics of $P^{(j)}_s$ used in \ci{K,DIKZ,DIK}.
However, we leave out all the Riemann-Hilbert analysis and just state
its results when needed. 
The details are given in the 3 mentioned publications.

First, we discuss the case of the sine-kernel.
In \ci{D}, Dyson found that 
\be\la{Dyson_as}
\ln P^{(1)}_s=-{s^2\over 2} -{1\over 4}\ln s+ c_0 +{a_1\over s}+ {a_2\over
s^2}+\cdots,\qquad s\to+\infty,
\ee
where 
\be 
c_0={1\over 12}\ln 2+ 3\ze'(-1).\la{c0}
\ee
Here $\ze(z)$ is the Riemann zeta-function. The constants $a_1$, $a_2$,
were also identified in \ci{D}. The first 2 leading terms in 
the expansion
(\ref{Dyson_as}) were found earlier by des Cloizeaux and Mehta \ci{dCM}.
The results in \ci{dCM} and \ci{D} were not fully rigorous.

The fact that the first leading term in (\ref{Dyson_as}) is correct
was proved in \ci{W2} by Widom. The full asymptotic expansion
of  $(d/ds)\ln P_s$ was obtained rigorously by Deift, Its, and Zhou in 
\ci{DIZ}. This result proves (\ref{Dyson_as}) up to the expression for 
$c_0$. The final step,
a proof that $c_0$ is given by (\ref{c0}), was carried out recently 
and in 3 variants: by Ehrhardt \ci{E}, by the author \ci{K}, and 
by Deift, Its, Zhou, and the author in \ci{DIKZ}. 
The methods of \ci{K} and \ci{DIKZ} are closely related and we will now
describe a ``hybrid'' approach based on these 2 papers.

For a function $f(\th)$ integrable over the unit 
circle, the Toeplitz determinant with symbol $f$ is given by 
the expression:
\be
D_n(f)=\det\left({1\over 2\pi}
\int_0^{2\pi}e^{-i(j-k)\th}f(\th)d\th\right)_{j,k=0}^{n-1}.
\ee
A Toeplitz determinant has the following two useful representations:
\be\label{multint}
D_{n}(f) = \frac{1}{(2\pi)^{n}n!}
\int_0^{2\pi}\cdots\int_0^{2\pi}
\prod_{1\leq j < k \leq n}|e^{i\theta_{j}} - e^{i\theta_{k}}|^{2}
\prod_{j=1}^n f(\th_j)d\theta_{j},
\ee
and
\be\la{Dchi}
D_n(f)=\prod_{j=0}^{n-1}\chi_k^{-2},
\ee
where $\chi_k$ are the leading coefficients of the polynomials
$\phi_k(z)=\chi_k z^k+\cdots$, $k=0,1,\dots$ orthogonal 
with weight $f(\th)$ on the unit circle. If $f(\th)$ is real
and nonnegative,
\be
\frac{1}{2\pi}
\int_0^{2\pi}\phi_k(e^{i\th})\overline{\phi_m(e^{i\th})}f(\th)d\th=
\delta_{km},\qquad k,m=0,1,\dots.
\ee

To obtain the asymptotics of the Fredholm determinants, we represent them
as double-scaling limits of Toeplitz (for the sine-kernel case)
and Hankel (for the Airy-kernel case, see below) determinants.
Let 
\be\la{f}
f(\th)\equiv f_\al(\th)=\cases{1,& $\al<\th<2\pi-\al$\cr
0,& otherwise}.
\ee
Now observe that 
\be\la{limT}
P^{(1)}_s=\lim_{n\to\infty}D_n(f_{2s/n}).
\ee
This fact is actually used in random matrix theory to obtain
the sine-kernel determinant; it was also used by Dyson in \ci{D}.
Note that if we could find the asymptotics of the polynomials orthogonal
with weight (\ref{f}), and in particular, the asymptotics of
$\chi_k$, $k\to\infty$, we would obtain by (\ref{limT}) and (\ref{Dchi})
part of (\ref{Dyson_as}) but not the constant $c_0$,
as the product of the first $\chi_0\chi_1\cdots\chi_{k_0}$ would
remain undetermined. However, this difficulty can be resolved.
We start with the following identity, which can be obtained from
(\ref{Dchi}):
\be
{d\over d\al}\ln D_n(f_\al)=
{n\over\pi}|\phi_n(e^{i\al},\al)|^2-{1\over\pi}\left\{
\phi_n(e^{-i\al},\al)e^{i\al}\phi_n'(e^{i\al},\al)+ \mathrm{
  c.c.}\right\},\qquad n=1,2,\dots\la{2det2}
\ee
where $\phi_k(z,\al)$ are the polynomials orthogonal 
w.r.t. $f_\al$ given by (\ref{f}), and $\phi_k'(z,\al)$ are their 
derivatives w.r.t. the variable $z$.
To use this identity, we now need to find the asymptotics of the 
polynomials appearing in the r.h.s. We do this by
solving the Riemann-Hilbert problem associated with these polynomials 
using a steepest descent approach of Deift and Zhou \ci{DZ}.
(Riemann-Hilbert formulation for orthogonal polynomials was first observed
in the case of orthogonality on the real line by Fokas, Its, and Kitaev in \ci{FIK}.) 
This step of the analysis is technically the most involved one.
If we substitute the results in the r.h.s. of (\ref{2det2}), we obtain
\be\la{diff}
{d\over d\al}\ln D_{n}(f_\al)=
  -{n^2\over 2}\tan{\al\over 2}-{1\over 8}\cot
{\al\over 2}+O\left({1\over n\sin^2(\al/2)}\right)
\ee
for all $n>s_0$ with some fixed $s_0$.
A crucial fact is that this expansion holds and the error term is 
uniform for ${2s_0\over n}\le\al<\pi$. 
We will now integrate this identity. 

First, we can obtain an expression for  $D_{n}(f_\al)$ as 
$\al\to\pi$ from below. Changing the variables $\th_j=\pi+(\pi-\al)x_j$
in (\ref{multint}) and expanding the integrand in $\pi-\al$, 
we obtain
\be\la{prom}
D_{n}(f_\alpha) = \frac{1}{(2\pi)^{n}n!}
\int_\al^{2\pi-\al}\cdots\int_\al^{2\pi-\al}
\prod_{1\leq j < k \leq n}|e^{i\theta_{j}} - e^{i\theta_{k}}|^{2}
\prod_{j=1}^n d\theta_{j}=
\frac{(\pi-\al)^{n^2}}{(2\pi)^n}A_n(1+O_n((\pi-\al)^2)),
\ee
as $\al\to\pi$ from below and $n$ is fixed.
Here
\be\la{A}
A_n={1\over n!}
\int_{-1}^{1}\cdots\int_{-1}^{1}
\prod_{1\leq j < k \leq n}(x_j - x_k)^{2}
\prod_{j=1}^n dx_j=2^{n^2}\prod_{k=0}^{n-1}
\frac{k!^3}{(n+k)!}
\ee
is a Selberg integral. Using its asymptotics as $n\to\infty$, we
obtain from (\ref{prom})
\be\la{alpi}
\ln D_n(f_\al)=n^2\ln\frac{\pi-\al}{2}-\frac{1}{4}\ln n+
c_0+\de_n+ O_n((\pi-\al)^2),\qquad \al\to\pi,
\ee
where $c_0$ is given by (\ref{c0}) and $\de_n\to 0$ as $n\to\infty$
($\de_n$ depends only on $n$).

Now we can integrate the identity (\ref{diff}) from $\al$ close to $\pi$
to $\al\ge 2s_0/n$ and use (\ref{alpi}) at the lower integration limit.
We thus obtain the following general formula:
\be\la{asf}
\ln D_n(\al)=n^2\ln\cos{\al\over 2}-{1\over
4}\ln\left(n\sin{\al\over 2}\right)+c_0+
O\left({1\over n\sin(\al/2)}\right)+\de_n,
\ee
for ${2s_0\over n}\le\al<\pi$, $n>s_0$, where $s_0$ is a
(large) positive constant.

Note that for a fixed $\al$, as $n\to\infty$ (\ref{asf}) 
reproduces a result 
of Widom \ci{Warc} for the asymptotics of a determinant on a fixed arc 
of the unit circle, which was used by Dyson to conjecture the value
of $c_0$ (\ref{c0}). 

Setting $\al=2s/n$, $s>s_0$ in (\ref{asf}), and letting $n\to\infty$,
we obtain by (\ref{limT})
\be
P^{(1)}_s=\lim_{n\to\infty}D_n(f_{2s/n})=
-{s^2\over 2} -{1\over 4}\ln s+ c_0 +O\left({1\over s}\right),
\ee
with $c_0$ given by (\ref{c0}). This, in particular, completes 
the proof for the constant term $c_0$ in (\ref{Dyson_as}).
Note that the present approach can be used to compute further terms 
in the asymptotic expansion.

We now turn our attention to the Airy-kernel determinant, $P^{(2)}_s$,
known as the Tracy-Widom distribution. In \ci{TW}, Tracy and Widom found
a connection of $P^{(2)}_s$ with the Hastings-McLeod solution of the 
Painlev\'e II equation, and also observed that
\be
\ln\det(I-K_s)=-{s^3\over 12}-{1\over 8}\ln s +\chi +
{b_3\over s^3}+{b_6\over s^6}+\cdots,
\qquad \mbox{as}\quad s\to +\infty,
\la{TW_as}
\ee
where the values of $b_3$, $b_6$, $\dots$ are extracted from the
asymptotics of the Hastings-McLeod solution, and
\be
\chi={1\over 24}\ln 2 +\ze'(-1).\la{const}
\ee
This value of $\chi$ was conjectured in \ci{TW} based on numerical 
evidence and by taking into account a similar expression for the constant
$c_0$ in (\ref{c0}). A proof was given by Deift, Its, and the author 
in \ci{DIK}, and another proof
by Baik, Buckingham, and DiFranco appeared in \ci{BBD}.
Here we discuss the approach used in \ci{DIK}, stressing its similarities 
to the method in the case of the sine-kernel described above.

For a function $w(x)$ integrable over the real half-line $(0,\infty)$, 
consider the Hankel determinant with symbol $w$:
\be
D_n^{H}(w)=\det\left(
\int_0^\infty x^{j+k}w(x)dx\right)_{j,k=0}^{n-1}.
\ee
Just as in the case of a Toeplitz determinant, the Hankel determinant
$D_n^{H}$ has the following two useful representations:
\be\label{multintH}
D_{n}^H(w) = \frac{1}{n!}
\int_0^{\infty}\cdots\int_0^{\infty}
\prod_{1\leq j < k \leq n}(x_j - x_k)^2
\prod_{j=1}^n w(x_j)dx_j,
\ee
and
\be\la{Dka}
D_n^H(w)=\prod_{j=0}^{n-1}\ka_k^{-2},
\ee
where $\ka_k$ are the leading coefficients of the polynomials
$p_k(x)=\ka_k x^k+\cdots$, $k=0,1,\dots$ orthogonal 
with weight $w(x)$ on the real half-line. If $w(x)$ is real and nonnegative,
\be
\int_0^{\infty}p_k(x)p_m(x)w(x)dx=
\delta_{km},\qquad k,m=0,1,\dots.
\ee

Let 
\be\la{w}
w(x)\equiv w_\al(x)=\cases{e^{-4xn},& $0<x<\al$\cr
0,& otherwise}.
\ee
With so defined $w_\al(x)$, the following analogue of (\ref{limT}) holds:
\be\la{limH}
P^{(2)}_s=\lim_{n\to\infty}\frac{D_n^{H}(w_{1-s/(2n)^{2/3}})}
{D_n^{H}(w_\infty)}.
\ee

Using (\ref{Dka}) we can obtain the following differential identity:
\be\la{idinterm}
{d\over d\al}\ln D_n^{H}(w_\al)={\ka_{n-1}(\al)\over \ka_n(\al)}e^{-4n\al}
(p'_{n}(\al,\al)p_{n-1}(\al,\al)-p_n(\al,\al)p'_{n-1}(\al,\al)),
\ee
where $p_k(x,\al)=\ka_k(\al)x^k+\cdots$ are the polynomials 
orthogonal on $(0,\al)$ with weight $w_\al(x)$, and 
the prime denotes differentiation w.r.t. the argument $x$.

A Riemann-Hilbert analysis of the polynomials 
$p_k(x,\al)$ as $k\to\infty$ produces
the asymptotic expression for the r.h.s. of (\ref{idinterm}), and 
we have
\be\la{di2}
{d\over d\al}\ln D_n^{H}(w_\al)
={n^2\over\al}(1-\al)^2+{\al\over 4(1-\al^2)}+
{1\over 1-\al}O\left({1\over n|1-\al|^{3/2}}\right).
\ee
This expansion 
holds uniformly in $\al\in(0,1-s_0/(2n)^{2/3}]$ for all $n>s_0^{3/2}/2$,
where $s_0$ is some (large) fixed number. 

To proceed as in the case of the sine-kernel, we estimate first 
$D_n^{H}(w_\al)$ for $\al\to 0$, where a series expansion can 
be written. This is done by an analysis of
(\ref{multintH}), and we obtain (cf. the derivation of (\ref{prom})):
\be\la{prom2}
  D_n^{H}(w_\al)={1\over n!}\int_0^\al\cdots\int_0^\al
\prod_{0\le i<j\le n-1}(x_i-x_j)^2
\prod_{j=0}^{n-1}e^{-4x_jn}dx_j=
\left({\al\over 2}\right)^{n^2}A_n(1+O_n(\al)),
\ee
as $\al\to 0$ from above and $n$ is fixed. The quantity $A_n$ is a Selberg
integral given by (\ref{A}).
Note that $D_n^{H}(w_\infty)$ is another Selberg integral:
\be\la{Dinf}
D_n^{H}(w_\infty)={1\over n!}\int_0^\infty\cdots\int_0^\infty
\prod_{0\le i<j\le n-1}(x_i-x_j)^2
\prod_{j=0}^{n-1}e^{-4x_jn}dx_j=(4n)^{-n^2}\prod_{k=0}^{n-1}k!^2.
\ee
(Both $A_n$ and $D_n^{H}(w_\infty)$ can also be computed using the formula
(\ref{Dka}) for the Legendre and Laguerre orthogonal polynomials, 
respectively.)

Using the asymptotics of $A_n$ and $D_n^{H}(w_\infty)$ for $n\to\infty$,
we conclude that (cf. (\ref{alpi}))
\be\la{Dal0}
\ln {D_n^H(w_\al)\over D_n^{H}(w_\infty)}=
\left({3\over 2}+\ln \al\right)n^2-{1\over 12}\ln{n\over 2}+
\ze'(-1)+\wt\de_n+O_n(\al),\qquad\al\to 0,
\ee
where $\wt\de_n$ depends on $n$ only, and $\wt\de_n\to 0$ as $n\to\infty$. 

Now we can integrate the identity (\ref{di2}) from $\al$ close to $0$
to $\al\le 1-s_0/(2n)^{2/3}$ and use (\ref{Dal0}) at the lower 
integration limit. We obtain for any $0<\al\le 1-s_0/(2n)^{2/3}$,
and any $n>s_0^{3/2}/2$ (cf. (\ref{asf})):
\be\eqalign{
\ln {D_n^H(w_\al)\over D_n^{H}(w_\infty)}=
n^2\left({3\over 2}+\ln\al-2\al+{\al^2\over 2}\right)-{1\over 12}\ln n -
{1\over 8}\ln (1-\al^2)+\\
{1\over 12}\ln2 +\ze'(-1)+
O\left(1\over n(1-\al)^{3/2}\right)+\wt\de_n.}\la{intD2}
\ee
Set here $\al=1-s/(2n)^{2/3}$, $s>s_0$, and let $n\to\infty$. By 
(\ref{limH}) we obtain
\be
P^{(2)}_s=\lim_{n\to\infty}\frac{D_n^{H}(w_{1-s/(2n)^{2/3}})}
{D_n^{H}(w_\infty)}=
-{s^3\over 12}-{1\over 8}\ln s + {1\over 24}\ln 2 +\ze'(-1)+O(s^{-3/2}),
\ee
which gives the first 3 terms of (\ref{TW_as}). Further terms can be 
obtained in this way as well.

Let us again stress that our approach \ci{K,DIKZ,DIK} to compute 
the asymptotics for
the above Fredholm determinants is based on approximating them with
Toeplitz and Hankel determinants. We then analyse the related systems
of orthogonal polynomials, and use identities for the logarithmic 
derivatives of Toeplitz and Hankel determinants. As a byproduct of 
this approach, we obtain the
asymptotics for the orthogonal polynomials (for weights $f_\al$ 
and $w_\al$).

\end{document}